\documentclass[final,3p,times]{elsarticle} 
\usepackage[colorlinks,
linkcolor=red,
anchorcolor=blue,
citecolor=green]{hyperref}

\usepackage{amssymb}
\usepackage{tabularx}
\usepackage{amsmath}
\newcommand\overbar[1]{\mkern 1.5mu\overline{\mkern-1.5mu#1\mkern-1.5mu}\mkern 1.5mu}
\usepackage{lineno}
\biboptions{authoryear}
\usepackage[figuresright]{rotating}

\usepackage{float}
\usepackage{graphicx}
\usepackage{epstopdf}
\journal{Journal of  Travel Research}
\usepackage{booktabs}
\usepackage{array}
\usepackage{threeparttable}
\usepackage{cleveref}
\usepackage{placeins}
\usepackage{caption}

\newcommand{\figref}[1]{\textcolor{blue}{Fig.~\ref{#1}}}
\newcommand{\tabref}[1]{\textcolor{blue}{Table~\ref{#1}}}
\newcommand{\Eqref}[1]{\textcolor{blue}{Eq.~(\ref{#1})}}
\newcounter{algorithm}

\newcommand{\algref}[1]{\textcolor{blue}{Algorithm~\ref{#1}}}
\usepackage{multirow}
\usepackage{array, caption, threeparttable}
\usepackage[font=small,labelfont=bf,labelsep=none]{caption}

\usepackage{bm}
\usepackage{pdflscape}
\usepackage{color}
\usepackage{xcolor}
\definecolor{gold}{rgb}{1.0, 0.84, 0.0}
\definecolor{silver}{rgb}{0.75, 0.75, 0.75}
\definecolor{copper}{rgb}{0.72, 0.45, 0.20}

\begin{document}
\captionsetup[figure]{labelfont={bf},labelformat={default},labelsep=period,name={Fig.}}
\begin{frontmatter}



\author[1]{Tingting Diao}
\address[1]{School of Computer, Electronics and Information,
Guangxi University, Nanning, Guangxi, 530004, China.}
\ead{dyxiscool@outlook.com}
\author[2]{Xinzhang Wu}
\address[2]{School of Electrical Engineering, Guangxi University, Nanning, Guangxi, 530004, China.}
\ead{dyxiscool@outlook.com}
\author[1]{Lina Yang}

\author[3]{Ling Xiao}\ead{xiaoling@cqupt.edu.cn}
\address[3]{School of Mathematics and Statistics, Xuzhou University of Technology, Xuzhou 221018, P.R. China}

\author[1]{Yunxuan Dong \corref{cor1}}
\cortext[cor1]{Corresponding author}

\title{A novel forecasting framework combining virtual samples and  enhanced Transformer models for tourism demand forecasting}

\begin{abstract}
Accurate tourism demand forecasting is hindered by limited historical data and complex spatiotemporal dependencies among tourist origins. A novel forecasting framework integrating virtual sample generation and a novel Transformer predictor addresses constraints arising from restricted data availability. A spatiotemporal GAN produces realistic virtual samples by dynamically modeling spatial correlations through a graph convolutional network, and an enhanced Transformer captures local patterns with causal convolutions and long-term dependencies with self-attention, eliminating autoregressive decoding. A joint training strategy refines virtual sample generation based on predictor feedback to maintain robust performance under data-scarce conditions. Experimental evaluations on real-world daily and monthly tourism demand datasets indicate a reduction in average MASE by 18.37\% compared to conventional Transformer-based models, demonstrating improved forecasting accuracy. The integration of adaptive spatiotemporal sample augmentation with a specialized Transformer can effectively address limited-data forecasting scenarios in tourism management.
\end{abstract}




\end{frontmatter}




\section{Introduction}
\label{introduction}

The accuracy of tourism demand forecasts affects the resource allocation, decision-making and service quality \citep{zhang2021tourism}. Spatiotemporal forecasting models show strong potential in analyzing fluctuations and seasonal trends by capturing the interactions between temporal and spatial regional trends. Spatiotemporal forecasting models based on deep learning can handle complex nonlinear spatiotemporal samples \citep{chen2023tourism}. However, deep learning models usually require massive samples to train their parameters. In addition, spatiotemporal features are complex and dynamic due to multiple spatial factors \citep{hamdi2022spatiotemporal}. With an insufficient sample and the increasing complexity of spatiotemporal features, existing research faces great difficulties in improving prediction accuracy and landing practical applications. Insufficient samples limit the model's ability\citep{li2025triangular}, while complex and dynamic spatial features further exacerbate the contradiction between sample demand and model performance \citep{wang2023multiscale}. Therefore, a dynamic framework capable of virtual sample generation and forecasting is necessary.

Generative Adversarial Networks (GANs) can generate virtual samples that are realistic to the original samples by implicitly learning the data distribution through adversarial training \citep{saxena2021generative}. The GAN structure usually combines a recurrent network structure and a multilayer convolutional structure to capture serial data features for time series data \citep{brophy2023generative}. Meanwhile, the Performance of GAN is further enhanced by conditional input and regularization based on gradient stabilization \citep{zhu2022partial}. However, focusing only on the temporal dimension cannot fully reflect the data's spatial features, limiting the virtual samples' quality. Therefore, the GAN model structure is extended to adapt to spatiotemporal data. Spatial effects in tourism demand express the complex dependencies between regions due to geographic proximity, seasons, and policies \citep{zhong2024analyzing,grossi2021seasonality,vojtko2022removing}. Such effects are dynamic, diverse, and challenging to capture with fixed rules. Furthermore, tourism demand has become more flexible with the spread of social media and is no longer limited to traditional geographic distances but is dynamically adjusted according to trends and preferences \citep{guizzardi2021big}. For example, certain attractions suddenly become popular destinations due to social media, which often cannot be captured by the fixed parameters in traditional models \citep{chen2023does,li2022you}. Therefore, static assumptions may lead to a loss of accuracy and flexibility in capturing dynamic changes in the model. In addition, combining various information forcibly into a static spatial weight matrix increases the model's complexity and may lead to feature mismatches and further undermine the model's generalization ability \citep{gao2023dynamic}. The GAN generator and discriminator incorporate Convolutional Neural Networks (CNNs) and Graph Convolutional Networks (GCNs) to capture the correlation of data at different spatial locations \citep{zhu2022spatiotemporal}. On the one hand, CNNs excel at dealing with local spatial features but rely on convolutional operations with fixed window sizes, which cannot effectively capture long-time dependencies \citep{jin2023spatio}. Such fixed sensory fields perform poorly for time series data with dynamic changes or significant long-term trends. On the other hand, existing research GCN structures depend on the static topology of the graph, which is difficult to adapt to real-time updated node and edge relationships \citep{jin2023spatio,wang2024integrated}.  Therefore, effectively incorporating dynamic spatial effects in GAN remains an urgent challenge.

The model mitigates data scarcity issues by augmenting the training set with virtual samples and enhances robustness in capturing spatiotemporal patterns.  However, ensuring that these enriched datasets translate into accurate forecasts requires a predictor capable of efficiently learning local and global temporal dependencies. Compared with recurrent architectures, Transformers avoid vanishing or exploding gradient issues and significantly improve training efficiency \citep{xu2024state}. However, Transformer performance is highly dependent on large-scale training data. Empirical studies indicate that Transformer models require tens of thousands of training samples to generalize effectively \citep{ganesan2021empirical}, which poses a challenge in the tourism forecasting domain where historical records are often sparse and limited. Furthermore, although self-attention mechanisms allow Transformers to model long-range dependencies effectively, they lack inductive biases that benefit time series forecasting, such as locality and recurrence \citep{shin2024attentive}. Unlike convolutional and recurrent architectures, standard Transformers rely solely on positional encoding to model time dependencies. This approach does not consider localized temporal variations, leading to suboptimal performance when dealing with short-term patterns and seasonal fluctuations \citep{wu2021autoformer}. Another major limitation of conventional Transformers in time series forecasting is their reliance on an autoregressive decoding mechanism. This step-by-step prediction strategy accumulates errors across forecasting horizons, especially when predicting long sequences \citep{zeng2023transformers}.  Additionally, Transformer models typically employ quadratic complexity due to their global attention computation, making them computationally expensive for long sequences. Various studies have proposed modifications to enhance Transformers for time series forecasting \citep{zhou2021informer,fu2024multivariate}. However, most approaches still require large datasets to achieve high performance, making them less suitable for data-limited scenarios such as tourism demand forecasting. Given the challenges posed by data scarcity and complex temporal patterns, designing a Transformer-based model that effectively captures both short-term and long-term dependencies while being robust under limited data conditions is necessary.

We propose a novel forecasting framework that integrates a spatiotemporal Generative Adversarial Network (GAN) with an enhanced Transformer predictor to address the challenges of data scarcity and complex temporal dependencies in tourism demand forecasting. The framework comprises two key components: the virtual sample generation module and the predictor module. The virtual sample generation module incorporates a spatiotemporal GAN to generate virtual samples by modeling spatial and temporal dependencies. This module employs a Graph Convolutional Network (GCN) and a Long Short-Term Memory Network (LSTM) to capture inter-regional correlations and temporal patterns. Unlike conventional GANs that rely on static spatial weight matrices, the proposed GCN dynamically computes the spatial weight matrix within a rolling window, ensuring real-time adaptability to evolving spatial interactions. While LSTM enhances temporal dependency modeling, it is complemented by the predictor to mitigate long-sequence forgetting loss. The predictor module uses an enhanced Transformer as a predictor designed to improve local feature extraction and long-term dependency modelling by integrating causal convolution with self-attention. Causal convolution first captures localized temporal dependencies before self-attention is applied to learn short-term variations effectively. Unlike traditional Transformer that rely on an autoregressive decoder, the enhanced Transformer eliminates the decoder and introduces a non-autoregressive global pooling mechanism, which mitigates error propagation and reduces computational complexity.  Furthermore, The framework employs a joint training strategy, where the predictor dynamically feeds back distribution deviations to the GAN, guiding the virtual sample generation process and improving consistency. \\
\indent This paper makes four contributions to tourism demand forecasting:
(1) This study integrates virtual sample generation with an enhanced Transformer for tourism demand forecasting under sample scarcity.
(2) A novel feature decoupling mechanism enhances spatial and temporal representation in small-sample scenarios.
(3) A spatiotemporal GAN-based framework improves virtual sample quality and diversity.

\begin{table}[htb]
\centering
\setlength{\abovecaptionskip}{0cm} 
\setlength{\belowcaptionskip}{-0.2cm} 
\caption{Research on virtual sample generation in recent five years}
\label{Research on virtual}
\begin{tabularx}{\textwidth}{p{3cm} X X X}
\toprule 
\textbf{Classification} & \textbf{Advantage} & \textbf{Disadvantage} & \textbf{Main model (Author Year)} \\
\midrule 
Information diffusion & Easy to implement; Remedy the insufficient information in sample & Sample dependency; Poor scalability and generalizability & Diffusion neural network \citep{yang2023diffusion}; Mega trend diffusion \citep{sivakumar2022synthetic} \\
Feature representation & Main for multi-dimensional sample; Maintain the core sample feature & Possibly introduce bias; Strong model dependence; High computational complexity; Exposure to loss of information & Neural network interpolation \citep{lin2023improved}; Based on singular value decomposition \citep{tian2021novel} \\
Deep learning & Strong learning ability & Model complexity; Consume large computing resources & GANs \citep{saxena2021generative}; Variational autoencoder \citep{huang2022boosting} \\
\bottomrule 
\end{tabularx}
\end{table}

\section{Literature review}
\label{Literature review}

\subsection{Spatiotemporal tourism demand forecasting}

Spatial effects reflect the influence of the geographic location and spatial distribution features on the interactions between variables, which are usually characterized by spatial spillover effects and spatial heterogeneity \citep{jiao2020forecasting}. Spatial effects are crucial in spatiotemporal tourism demand forecasting because tourism demand is not only driven by time-series changes but also significantly depends on geographical proximity and interconnections between regions. \citet{marrocu2013different} pointed out that the movement of tourists is affected not only by the geographic distance and specific features of origins and destinations but also by the neighboring locations of origins and destinations. Spatial spillover effects are reflected in the flow of tourists between areas and the spread of tourist attractions \citep{kim2021spatial}. In contrast, spatial heterogeneity is reflected in the heterogeneous distribution of tourism demand in different areas due to the economic \citep{brida2020empirical}, cultural \citep{vergori2020cultural} and policy \citep{collins2021national}. Therefore, spatiotemporal tourism demand forecasting models need to capture both temporal and spatial interactions to fully reveal the complex patterns of tourism demand and improve forecasting accuracy.\\
 \indent There are two main models for Spatiotemporal tourism demand forecasting: spatial econometric and deep learning models. \citet{deng2011modelling} used an Anisotropic Dynamic Spatial Lag Panel Origin-Destination Tourism Flow model to model the Australian domestic and international inbound tourism. Specifically, their study addressed the traditional limitations in a single cross-sectional context. Further exploring spatial-temporal dynamics, \citet{yang2019spatial} applied the Dynamic Spatial Panel models and Spatiotemporal Autoregressive Moving Average models with varying specifications of spatial weighting matrices to forecast the inbound tourism demand in 29 Chinese provincial regions. Their findings demonstrated that spatial-temporal models yielded lower average forecasting errors than a-spatial models, particularly in regions with strong local spatial associations. Building on spatial effects in tourism forecasting, \citet{jiao2020forecasting} employed the Global and Local Spatiotemporal Autoregressive models to forecast the tourist arrivals for 37 European countries. Although spatial econometric models such as the Dynamic Spatial Panel and Spatiotemporal Autoregressive models have made some achievements in capturing spatial effects, traditional spatial econometric models fail to handle complex nonlinear features. Researchers began to explore deep learning models for spatiotemporal tourism demand forecasting.\\
\indent \citet{li2022tourism} used a graph convolution network to extract spatial effects. The spatial weight matrix incorporates information such as proximity, demand recognition, and arrivals. \citet{zhou2023graph} proposed a spatiotemporal learning framework based on graph attention deep. The spatial weight matrix incorporates holidays, weather, and distance information. In the study above, the spatial weight matrix fuses multiple information to capture the time series features as comprehensively as possible. Multiple weight matrices improve representation capabilities but significantly increase computational complexity and may introduce redundant information. In addition, most of the existing spatial weight matrices are based on geographic distance and static correlation coefficients. A weight matrix based on geographic distance is the basic way to quantify inter-regional dependence through proximity. However, the models based on geographic distance cannot reflect the role of non-geographic factors such as economic ties or culture \citep{liu2023exploring}. Static correlation-based weight matrices utilize historical samples to calculate inter-regional correlations. Although they can capture certain dynamic relationships, they are susceptible to sample noise and challenging to adapt to real-time changes \citep{chen2023tourism,shen2024short}. \\
\indent Multiple spatial weights increase the complexity of the model and may require extensive tuning. Predefined spatial weight matrices effectively capture basic spatial dependencies and provide structure for inter-regional modeling. However, the spatial relationships of virtual samples often may not align with a preset matrix after data augmentation. Therefore, dynamically learning spatial dependencies without relying on fixed matrices is essential.

\subsection{The GANs for virtual sample generation}
Virtual sample generation models generate virtual samples by learning the features carried by samples and combining the prior knowledge. Virtual sample generation can be divided into three categories as shown in \tabref{Research on virtual}. Virtual sample generation based on deep learning has become one of the most popular methods in recent years. Among deep learning methods, GANs play an important role \citep{wang2017generative}. Researchers study and refine the GAN-based models, loss functions, and optimization methods to meet diverse application needs. \citet{radford2015unsupervised} developed the Deep Convolutional Generative Adversarial network (DCGAN) in the following year. The initial GAN-based models suffered from the gradient disappearance problem, which hindered the generator's ability to learn efficient distributions.  \citet{arjovsky2017wasserstein} proposed the Wasserstein Generative Adversarial Network (WGAN), which employed the Earth-Mover distance as a new optimization objective to measure distribution differences more effectively. \citet{gulrajani2017improved} proposed WGAN-GP, adding a gradient penalty to WGAN to further improve training stability and the overall quality of generated samples significantly.

The quality of generated virtual samples has improved, which remains challenging when handling tasks with temporal and spatial dependencies. The challenge is that GAN-based models emphasize capturing temporal dependencies while overlooking spatial dependencies. \citet{mogren2016c} proposed the first GAN-based models designed to generate continuous sequences  (C-RNN-GAN) by incorporating recurrent structures to preserve temporal dependencies. \citet{esteban2017real} proposed a Recurrent Condition Generative Adversarial Network (RCGAN), which improved on C-RNN-GAN by removing dependence on previous outputs and incorporating additional information to guide sample generation. \citet{yoon2019time} introduced self-supervised learning and encoder-decoder structures to learn patterns in time series and proposed the Time Series Generative Adversarial Networks (TimeGAN). \citep{brophy2023generative} comprehensively summarized the latest applications and improvements of GAN-based models in generating time series samples.
However, the GAN-based models mentioned do not consider spatial effects and mainly focus on modeling time-dependent features effectively and efficiently.

Although existing GAN-based models rarely focus on the potential spatial effects of time series, extensive studies on spatiotemporal events and trajectory data are being conducted. For example, \citet{yu2020extracting} applied CGAN with LSTM to capture spatial and temporal changes in taxi hotspots. \citet{jin2019crime} extracted latent variables from multi-channel graph datasets using a combination of CNN and variational autocoder methods. \citet{liu2020col} utilized a CNN-based network as a trajectory discriminator. Unlike other GAN-based trajectory prediction methods, the proposed discriminator can classify each segment. \citet {huang2023robust} proposed a spatiotemporal generative adversarial input network to deal with the spatiotemporal dependencies and missing conditions. \citet {hu2023multiload} proposed a multi-load generative adversarial network for simultaneously generating load profiles considering spatiotemporal dependencies. In addition, \citet{lei2019gcn} proposed GCN-GAN to predict trajectories in weighted dynamic networks. However, \citet{lei2019gcn} did not consider the effect of exogenous variables on the trajectory. Furthermore, no current GAN considers exogenous variables and dynamic spatial effects when generating virtual samples, especially in tourist flow. Our study builds on these insights to address this gap by proposing a enhanced approach integrating exogenous variables and dynamic spatial effects into the GAN framework.\\
\indent \subsection{Transformer-based models as predictor}
The Transformer model has demonstrated remarkable capabilities in capturing dependencies across different time steps within a sequence through self-attention mechanisms. Compared with traditional Recurrent Neural Networks, the Transformer addresses the problem of vanishing or exploding gradients often encountered when handling long-distance dependencies. Furthermore, the self-attention mechanism enables efficient modeling of both short-term and long-term dependencies, making it well-suited for time series forecasting.\\
\indent Despite its advantages, the application of the Transformer in tourism demand forecasting remains limited. This is primarily due to the relatively small size of available datasets in the tourism domain, which makes it difficult to fully exploit the potential of the Transformer's data-intensive architecture. Additionally, collecting comprehensive and high-quality tourism data is a time-consuming process, further hindering its application in this field.\\
\indent To our knowledge, only four studies have explored attention mechanisms in tourism demand forecasting. \citet{zheng2021multi} used LSTM and an attention mechanism for tourism demand forecasting; \citet{zhou2023graph} proposed an attention network combined with a graph structure; \citet{dong2023time} introduced a guided-attention model to identify seasonal and non-smooth features; \citet{zhang2023crossformer} proposed a spatiotemporal transformer network that integrates temporal and spatial transformer modules with a spatiotemporal fusion module to capture dynamic dependencies and correlations. While these approaches leverage attention mechanisms for improved forecasting, they often require complex spatial dependency structures or large datasets, which can limit their scalability and practical applicability.

\section{Proposed method}
\label{Proposed method}

\subsection{Virtual sample generation module for forecasting framework}
Given a noise vector $\bm z$ and a real sample $\bm x$, the generator utilizes the function $G(\bm z)$ to transform unstructured noise into virtual samples. The discriminator evaluates both virtual and real samples, returning $D(G(\bm z))$ and $D(\bm x)$, which estimate whether a sample originates from the real or generated distribution. The generator is trained to produce indistinguishable samples from real ones, forcing the discriminator to fail in distinguishing between the two distributions. The derivative of the discriminator's output concerning its input is propagated back to the generator. In the standard GAN framework, the discriminator is trained using a binary classification objective, where real samples are encouraged to have a high probability. In contrast, virtual samples are pushed towards a low probability. However, this formulation often suffers from unstable training and mode collapse. We adopt an improved adversarial training strategy to address these challenges that replace the traditional binary classification objective with a real-valued scoring mechanism. Instead of classifying samples as real or fake, the discriminator assigns a continuous score to each sample, reflecting its relative authenticity. Additionally, we introduce a gradient penalty term, which enforces smoothness by penalizing deviations of the discriminator's gradient norm from 1. The final objective function is formulated as:
\begin{align} \label{wgan_total}
\min_{G} \max_{D} \bm L_{\text{total}} = \bm L_D + \lambda_1 \bm L_G + \lambda_2 \bm L_{GP},
\end{align}
\begin{align} \label{wgan_discriminator}
\bm L_D = - \left( \mathbb{E} [D(\bm x)] - \mathbb{E} [D(G(\bm z))] \right),
\end{align}
\begin{align} \label{wgan_generator}
\bm L_G = - \mathbb{E} [D(G(\bm z))],
\end{align}
\begin{align} \label{gp}
\bm L_{GP} = \mathbb{E} \left[ (\|\nabla_{\hat{\bm x}} D(\hat{\bm x})\|_2 - 1)^2 \right].
\end{align}
where $\lambda_1$ and $\lambda_2$ are weight factors balancing the generator's loss, discriminator's loss, and gradient penalty, respectively. $G$ and $D$ denote the generator and discriminator, respectively, while $\mathbb{E}_x$ and $\mathbb{E}_z$ represent the expected values over real and generated samples. The gradient penalty term $\bm L_{GP}$ enforces constraints on the discriminator by penalizing deviations from the desired gradient norm, ensuring stable convergence.

We construct a fully connected graph between countries or regions and compute a dynamic spatial weight matrix using correlation coefficients to characterize the spatial dependence of the graph. Unlike the traditional method in which the correlation coefficients are computed for all time steps to generate a static spatial weight matrix, we introduce a rolling window mechanism to dynamically compute the correlation coefficients from the historical time step $q$ and generate or forecast time step $Q$. The spatial weight matrix can be updated to capture the time-varying spatial relationships more accurately. At time $t$, the tourism feature vector of the $n$th country or region is denoted as $\bm{x}_n(t)$, where $t \in [1, T]$, $n \in [1, N]$, $T$ is the maximum value of the time step, and $N$ is the total number of countries or regions. The $N$ countries or regions feature matrix is denoted as $\bm{X}(t) = {\bm{x}_1(t), \bm{x}_2(t), \ldots, \bm{x}_N(t)}$, where $\bm{X}(t) \in \mathbb{R}^{N \times 1}$. Using the $q$ historical time steps of the rolling window, the feature matrix of $N$ countries or regions as $\bm{X}{(t, q)} = (\bm{X}(t-q+1), \bm{X}(t-q), \ldots, \bm{X}(t)) \in \mathbb{R}^{Q \times N \times 1} $. Our goal is to generate virtual samples $\bm{\hat{X}}(t, q) \in \mathbb{R}^{N \times Q}$.

 For each historical time step $q$, we dynamically calculate the correlation coefficient $a_{ij}$ between the $i$th country or region and the $j$th country or region through the feature matrix within the rolling window as follows:
\begin{align}
a_{ij} = \frac{\sum_{u=u_0}^{t} ({x}_{i}^{(1)}(u) - \overbar{{\bm X}}_i^{(1)}(t, q))({x}_{j}^{(1)}(u) - \overbar{{\bm X}}_j^{(1)}(t, q))}{\sqrt{\sum_{u=u_0}^{t} ({x}_{i}^{(1)}(u) - \overbar{{\bm X}}_i^{(1)}(t, q))^2} \sqrt{\sum_{u=u_0}^{t} ({x}_{j}^{(1)}(u) - \overbar{{\bm X}}_j^{(1)}(t, q))^2}},
\end{align}
where $u_0=t-q+1$. ${x}_{i}^{(1)}(u)$ and ${x}_{j}^{(1)}(u)$ respectively represent tourists demand at time $u$ in the $i$th and $j$th country or region. $\overbar{{\bm X}}_i^{(1)}(t, q)$ and $\overbar{{\bm X}}_j^{(1)}(t, q)$ respectively represent the average tourists demand in the $i$th and $j$th country or region from time $t-q+1$ to $t$. $\bm{A}{(t, q)}$ represents spatial weight matrix for all regions from time $t-q+1$ to $t$, where $\bm{A}{(t, q)} = (a_{ij})_{N \times N}$. $\bm{A}{(t, q)} \in \mathbb{R}^{Q \times N \times N}$. The real features may be altered when the correlation coefficient matrix is directly used for the convolution calculation. We normalize spatial weight matrix $\bm A{(t, q)}$ and denoted as $\tilde {\bm{A}}{(t, q)}$.

The spatiotemporal GAN captures spatial features and gains a series with spatial features after two layers of convolution operations. The convolution operations for feature extraction as follows:
\begin{align}
\bm{H}^{(L)}=  \tilde {\bm{A}}{(t, q)} \text{ReLU} \left( \tilde {\bm{A}}{(t, q)} (\bm X{(t, q)}) \bm W^{(0)} \right)  \bm W^{(1)},
\end{align}
where $ \bm {H}^(L) \in \bm{R}^{Q \times N \times F_2}$. ReLU$ (\cdot)$ represents the activation function. $\bm W^{(0)}$ and $\bm W^{(1)}$ represent weight matrices. $\bm W^{(0)} \in \mathbb{R}^{ Q \times 1\times F_1}$, $\bm W^{(1)} \in \mathbb{R}^{Q \times  F_1\times F_2}$. $F_1$ and $F_2$ represent the output feature dimensions of convolution operations' first and second layers.

After obtaining the spatial feature representation $ \bm H^{(L)}$ for each time step $t$, LSTM is used to capture the temporal dependencies. In the LSTM, $\bm {H}^{(L)}$ and previous temporal features $\bm h_{t-1}$ are used to compute memory unit $\bm{c}_t$.
\begin{align}
\bm{c}_t = \bm{f}_t \odot \bm{c}_{t-1} + \bm{i}_t \odot \tanh(\bm{H}_t^{(L)} \bm{W}_{xc} + \bm{h}_{t-1} \bm{W}_{hc} + \bm{b}_c),
\end{align}
\begin{align}
\bm h_t = \bm o_t \odot \tanh(\bm c_t)= \text{LSTM}(\bm H_t^{(L)}, \bm h_{t-1})),
\end{align}
\begin{align}
\bm{\hat{X}}({t,q}) = \bm h_t\cdot \bm W_g + \bm b_g,
\end{align}
where $\bm{H}_t^{(L)} \in \bm{H}^{(L)} = \{\bm{H}_1^{(L)}, \bm{H}_2^{(L)}, \dots, \bm{H}_Q^{(L)}\}$. $\bm{h}_{t-1}$ is the hidden state vector in the LSTM, which contains history features captured from time 0 to $t$-1, $\bm{h}_{t-1} \in \mathbb{R}^{ N \times d}$. $d$ represents hidden cell num.  $\bm{i}_t$, $\bm{f}_t$ and $\bm{o}_t$ represent the input gate, forget gate and output gate. $\bm W_{xc}$, $\bm W_{hc} $ and $\bm{W}_g$ are the weight matrices. $\bm W_{xc} \in \mathbb{R}^{F_2 \times d}$ , $W_{hc} \in \mathbb{R}^{d \times d}$ , $\bm{W}_g \in \mathbb{R}^{d \times Q}$. $\bm{b}_c$ and $\bm{b}_g$ are bias vectors.
$\bm{b}_c \in \mathbb{R}^{d}$, $\bm{b}_g \in \mathbb{R}^{Q}$. The $\tanh(\cdot)$ represents the activation function. $\odot$ represents element-wise multiplication. A detailed analysis is required due to the differing inputs and outputs of the generator and discriminator in the spatiotemporal GAN. The algorithm details of the spatiotemporal GAN model are shown \algref{Algorithm 1}.

\begin{table}[H]
    \centering
    \refstepcounter{algorithm}\label{Algorithm 1}
    \begin{tabularx}{\columnwidth}{X}
        \toprule
        \textbf{Algorithm 1: The Proposed Spatiotemporal GAN}\\
        \midrule
        \textbf{1. Initialize Parameters:} \\
        - Learning rates $\alpha_G$, $\alpha_D$, $\alpha_T$ of generator, discriminator\\, and predictor after pre-training\\
        - Total iterations $\text{Max\_iters}$ \\
        \textbf{2. Compute Adjacency Matrices:}
        $\tilde {\bm{A}}{(t, q)}$ \\
        \textbf{3. Generator $G(\bm{X}(t,q), \tilde {\bm{A}}{(t, q)}$ Process:} \\
        \textbf{Input:}$\bm{z}(t,q) \in \mathbb{R}^{Q \times N \times 1}$,  $\tilde {\bm{A}}{(t, q)}$ \\
        3.1 Apply GCN layers: $\bm{H}^{(L)} \in \mathbb{R}^{Q \times N \times F}$  \\
        3.2 Apply LSTM with $\bm H^{(L)}$: \\
        \hspace{1em} for each time step $t$ in range $(1, Q + 1)$: \\
        \hspace{2em} $ \bm h_t, \bm c_t \leftarrow \text{LSTM}(\bm H_t^{(L)}, \bm h_{t-1}, \bm c_{t-1})$ \\
        3.3 Generate virtual samples: $\bm{\hat{X}}(t,q) \leftarrow \bm h_t\cdot \bm W_g + \bm b_g$ \\
        \textbf{4. Discriminator $D(\bm{X}(t, q), \tilde {\bm{A}}{(t, q)}$ Process:} \\
        \textbf{Input:}$\bm{X}(t,q) \in \mathbb{R}^{Q \times N \times 1}$, $\tilde {\bm{A}}{(t, q)}$ \\
        4.1 Apply GCN layers to obtain $\bm H_{\text{disc}}^{(L)} \in \mathbb{R}^{Q \times N \times F}$ \\
        4.2 Apply LSTM: Initialize $\bm h_{\text{disc}, 0}$, $\bm c_{\text{disc}, 0}$ \\
        \hspace{1em} for each time step $t$ in range $(1, Q + 1)$: \\
        \hspace{2em} $\bm h_{\text{disc}, t}, \bm c_{\text{disc}, t} \leftarrow \text{LSTM}(\bm H_{\text{disc}, t}^{(L)}, \bm h_{\text{disc}, t-1}, \bm c_{\text{disc}, t-1})$ \\
        4.3 Output: $D(\bm{X}(t, q), \tilde {\bm{A}}{(t, q)} \leftarrow \bm h_{\text{disc}, t} \cdot \bm W_{\text{disc}} + \bm b_{\text{disc}}$ \\
        \textbf{5. Training Spatiotemporal GAN:} \\
        Initialize iteration counter $\text{iter} = 0$ \\
        \textbf{while} $\text{iter} < \text{Max\_iters}$ \textbf{do} \\
        \hspace{1em} \textbf{5.1 Update Discriminator (D):} \\
        \hspace{2em} Generate samples $\bm{\hat{X}}(t,q) \leftarrow G(\bm{X}(t,q), \tilde {\bm{A}}{(t, q)})$ \\
        \hspace{2em} Compute Discriminator loss: \\
        \hspace{2em} $L_D = -\mathbb{E} [D(\bm{X}(t, q), \tilde {\bm{A}}{(t, q)}] - \mathbb{E} [D(\bm{\hat{X}}(t,q), \tilde {\bm{A}}{(t, q)}]$ \\
        \hspace{2em} Compute Gradient Penalty: \\
        \hspace{2em} $L_{GP} = \mathbb{E} \left[ \left( \left\| \nabla_{\hat{\bm X}} D(\hat{\bm{X}}(t,q), \tilde {\bm{A}}{(t, q)} \right\|_2 - 1 \right)^2 \right]$ \\
        \hspace{2em} Total Discriminator Loss: $\tilde{L}_D = L_D + \lambda_{GP} L_{GP}$ \\
        \hspace{2em} Backpropagate and update: \\
        \hspace{2em} $\theta_D \leftarrow \theta_D - \alpha_D \nabla (\tilde{L}_D)$ \\
        \hspace{1em} \textbf{5.2 Update Generator (G) and predictor:} \\
        \hspace{2em}  Generator loss:$L_G = -\mathbb{E}[D(\bm{\hat{X}}(t,q), \tilde {\bm{A}}{(t, q)}] \cdot \lambda_{\text{gan}}$ \\
        \hspace{2em} predictor Loss: $L_{\text{pred}} = \mathbb{E}[\| \text{predictor}(\bm{\hat{X}}(t,q)- \bm{X}(t, q) \|_2^2] \cdot \lambda_{\text{pred}}$ \\
        \hspace{2em} Total Loss: $\tilde{L}_G = L_G + L_{\text{pred}}$ \\
        \hspace{2em} Backpropagate and update: \\
        \hspace{2em} $\theta_G \leftarrow \theta_G - \alpha_G \nabla (\tilde{L}_G)$ \\
        \hspace{2em} $\theta_T \leftarrow \theta_T - \alpha_T \nabla (\tilde{L}_G)$ \\
        \hspace{1em} \textbf{5.3 Update iteration counter:} $\text{iter} \leftarrow \text{iter} + 1$ \\
        \textbf{end while} \\
        \midrule
    \end{tabularx}
\end{table}

\subsection{Predictor module for forecasting framework}
We use an enhanced Transformer-based model as a predictor that incorporates causal convolution for local feature extraction, positional encoding for improved temporal representation, and a non-autoregressive forecasting approach via global pooling. These enhancements mitigate the limitations of traditional Transformer models by improving feature extraction, computational efficiency, and interpretability in time series forecasting.

The key contributions of our proposed model are as follows:
\begin{itemize}
    \item 	Local Feature Extraction via Causal Convolution: We introduce a causal convolution module to capture short-term temporal dependencies before applying self-attention, ensuring that the model effectively learns localized patterns without information leakage.
    \item 	Enhanced Temporal Encoding: Unlike standard Transformer models that solely rely on position encoding, we combine causal convolution with sinusoidal positional encoding to provide a more comprehensive representation of sequential dependencies.
    \item 	Non-Autoregressive Forecasting with Global Pooling: Instead of employing an autoregressive decoder, we utilize a global pooling mechanism to aggregate sequence-wide contextual information, reducing error accumulation and improving prediction stability.
\end{itemize}

\subsubsection{Model overview}
\begin{figure}[H]\centering
	\includegraphics[width=\linewidth]{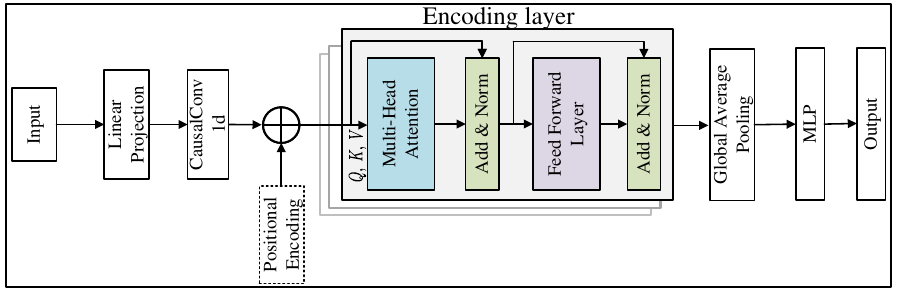}
	\caption{The proposed predictor module for forecasting framework.}
    \label{Transformer}
\end{figure}

The predictor consists of an embedding layer, a causal convolution module, a stack of Transformer encoder layers, a global pooling operation, and a prediction head. The structure is illustrated in \figref{Transformer}.

\subsubsection{Causal convolution for local feature extraction}
Unlike the standard Transformer, which solely relies on self-attention, we introduce a causal convolution layer to extract short-term temporal dependencies before applying the self-attention mechanism. Given a time series sample $\bm X_i(t, q) \in \mathbb{R}^{q \times 1}$ for region $i$, the embedding layer first transforms the input into a higher-dimensional space:
\begin{align}
    \bm{E}_i = \bm X_i(t, q) \bm W_E + \bm b_E,
\end{align}
where $\bm W_E \in \mathbb{R}^{1 \times d}$ and $\bm b_E \in \mathbb{R}^{d}$ represent learnable embedding parameters.

The embedded features are then processed by a causal convolution layer:
\begin{align}
    \bm{C}_i = \text{CausalConv1D}(\bm{E}_i, \bm W_C),
\end{align}
where $\bm W_C$ denotes the convolutional kernel parameters. This operation enhances local temporal representations while preserving causality by ensuring that future information is not used.

\subsubsection{Positional encoding and transformer encoder}
To capture long-term dependencies, we apply sinusoidal positional encoding to the output of the causal convolution:
\begin{align}
    \bm{P}_i = \text{PositionalEncoding}(\bm{C}_i).
\end{align}
The encoded sequence is then passed through multiple Transformer encoder layers. Each encoder layer consists of a multi-head self-attention mechanism, followed by a feedforward network with residual connections and layer normalization. The self-attention mechanism calculates the attention scores for each head $j$ as:
\begin{align}
\bm{Self}_i^{(j)} = \text{softmax}\left( \frac{\bm Q_i^{(j)} (\bm K_i^{(j)})^T}{\sqrt{d_k}} \right) \bm V_i^{(j)},
\end{align}
where $\bm Q_i^{(j)}$, $\bm K_i^{(j)}$, and $\bm V_i^{(j)}$ are derived from the input sequence:
\begin{align}
\bm Q_i^{(j)} = \bm{P}_i \bm W_Q^{(j)}, \quad \bm K_i^{(j)} = \bm{P}_i \bm W_K^{(j)}, \quad \bm V_i^{(j)} = \bm{P}_i \bm W_V^{(j)}.
\end{align}
The outputs of different attention heads are concatenated and projected:
\begin{align}
\bm{MultiSelf}_i = \text{Concat}(\bm{Self}_i^{(1)},\dots, \bm{Self}_i^{(h)})\bm W_O.
\end{align}
The final output of the Transformer encoder is obtained after applying residual connections and layer normalization:
\begin{align}
\bm{H}_i = \text{LayerNorm}(\bm{P}_i + \bm{MultiSelf}_i).
\end{align}

\subsubsection{Global pooling and prediction head}
Instead of employing an autoregressive decoder, we adopt a global average pooling mechanism to aggregate information across the entire sequence.The pooled features are then passed through a feedforward network to refine representations:
\begin{align}
\bm{H}_{\text{pool}} = \frac{1}{q} \sum_{t=1}^{q} \bm{H}_i(t),
\end{align}
\begin{align}
\bm{FFN}(\bm H_{\text{pool}}) = \text{ReLU}(\bm H_{\text{pool}} \bm W_1 + \bm b_1) \bm W_2 + \bm b_2.
\end{align}
where $\bm{H}_{\text{pool}}$ represents the globally pooled feature vector.

Finally, the output is projected  using linear transformation:
\begin{align}
\bm{\tilde{X}}(t+p) = \bm{FFN}(\bm H_{\text{pool}}) \bm W_{out} + \bm b_{out}.
\end{align}

\subsection{Joint training predictor and virtual sample generation module}
\begin{figure}[H]\centering
	\includegraphics[width=\linewidth]{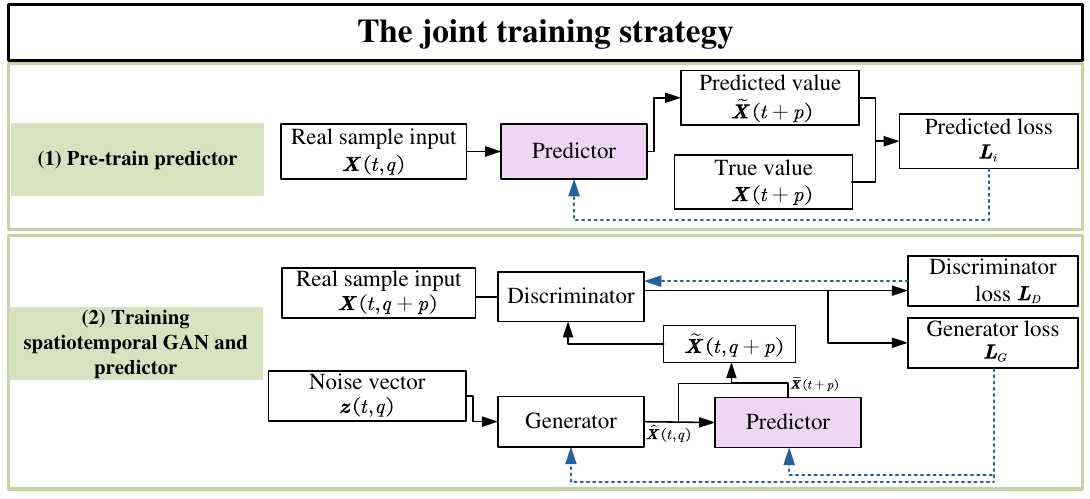}
	\caption{The joint training of the proposed predictor. The strategy is divided into two parts: (1) Pre-training the predictor model. (2) Training the spatiotemporal GAN and predictor model.}
    \label{model}
\end{figure}

Because of limited samples, we use joint training of the spatiotemporal GAN and predictor to improve virtual sample quality and predicted accuracy. The architecture of joint training is shown in \figref{model}. This strategy is divided into two steps: pre-training the predictor and joint training the spatiotemporal GAN and predictor.

\noindent \textbf{(1) Pre-training predictor:} pre-training the predictor is shown in Figure \figref{model}(1). The input sample $\bm {{X}}(t,q)$ is processed by the predictor to obtain the predicted value $\bm {\tilde{X}}({t+p})$. The predicted value is compared with the true value $\bm {{X}}({t+p})$, and the prediction loss $L_i$ is calculated between the predicted value and the true value. This process uses Mean Square Error (MSE) loss to optimize the model.

\noindent \textbf{(2) Training spatiotemporal GAN and predictor:} The traditional method usually directly combines the virtual sample and the real sample into the predictor. The traditional method often results in error accumulation and does not allow dynamic tuning of model parameters. We integrate the predictor into the spatiotemporal GAN architecture. The Generator use noise vector $\bm{z}(t, q)$ and a conditional vector $\bm{y}(t, q)$ to generate virtual sample $\bm{\hat{X}}(t, q)$. The predictor further processes virtual sample to forecast $p$ steps $\bm{\overline{X}}(t + p)$. The discriminator uses the virtual output $\bm{\tilde{X}}(t, q + p)$, the real sample $\bm{X}(t, q + p)$, and the conditional vector $\bm{y}(t, q + p)$ as input to determine whether the sample is real or generated.
The pre-trained predictor uses the virtual samples to make predictions. The discriminator calculates the loss $L_D$ by comparing the real samples with the predicted and virtual samples. We optimize the model parameters by back propagation. The generator loss $L_G$ regulates the generator and predictor.

\section{Empirical study}
\label{Empirical study}

\subsection{Sample analysis}
\begin{figure}[htb]\centering \includegraphics[width=\linewidth]{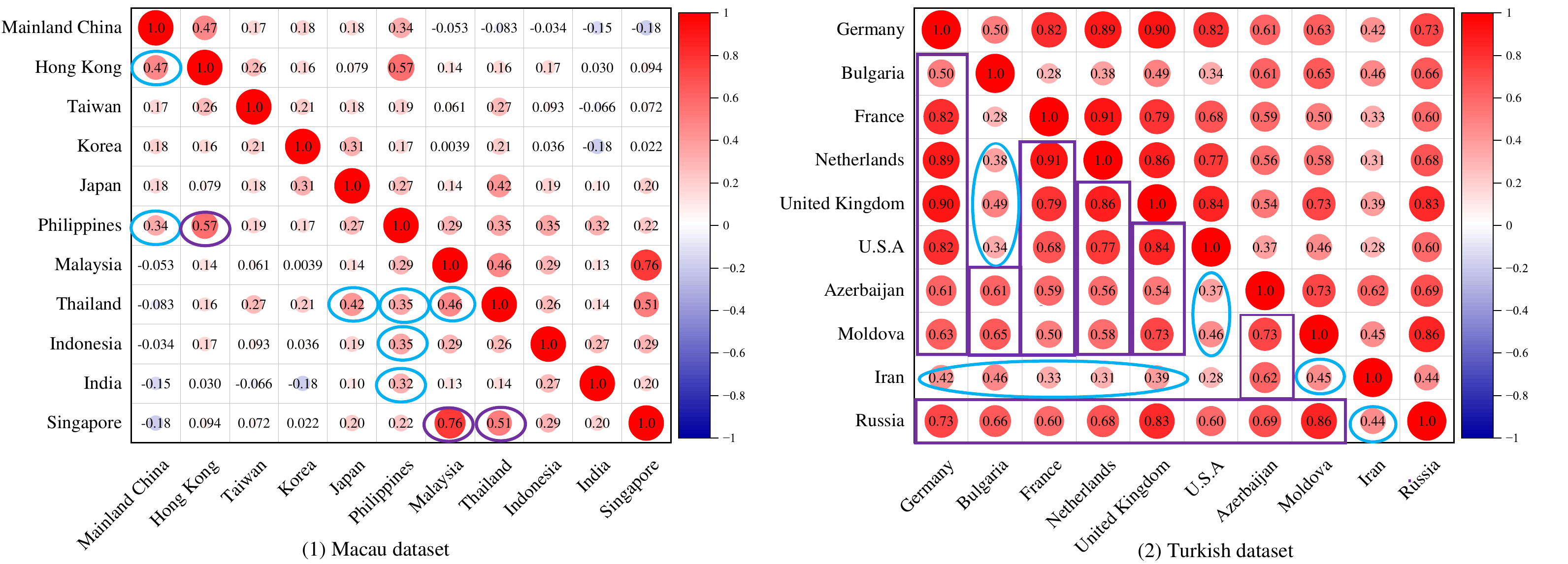}
\caption{Heat map of correlation among countries or regions (The purple border highlights strong correlations (\(|r| \geq 0.5\)), while the blue border indicates moderate correlations (\(0.3 \leq |r| < 0.5\)).).}
\label{The correlation}
\end{figure}

We use two datasets to validate the effectiveness of the proposed model. One dataset is the daily tourist arrivals to Macau from 11 regions or countries from 1 January 2017 to 30 May 2019. The other dataset is the monthly tourist arrivals to Turkey from 10 countries from April 2008 to March 2020. Macau and Turkey datasets complement each other by covering different temporal granularities and distinct geographical contexts (Asia and Europe), which comprehensively evaluate the model's ability to capture diverse spatiotemporal patterns. We plot a heat map of the Pearson correlation coefficients based on the average tourism arrivals across all years as shown in \figref{The correlation}. The analysis in \figref{The correlation} highlights that the diverse spatial effects create significant variability in correlation patterns. These results underscore the importance of employing a dynamic spatial weight matrix and adaptive learning strategies to capture and model these complex relationships effectively.

\noindent \textbf{(1) Macau dataset}

The heat map shows a moderate correlation of tourists between Mainland China and Hong Kong, which can be attributed to their geographic proximity to Macau, the convenience of transportation (e.g., the opening of the Hong Kong-Zhuhai-Macau Bridge and the high-speed railway), and a language and cultural commonality. In addition, the moderate correlation between the Philippines and Hong Kong can be attributed to historical and cultural ties (Catholicism and Western architectural influences) and the convenience of Philippine tourists transiting to Macau via Hong Kong. The high correlation between Southeast Asian countries such as Malaysia and Singapore reflects their shared interest in Macau tourism, which may be influenced by a combination of similar cultural attractions (e.g. Chinese culture and cuisine) and marketing strategies. However, for countries that are geographically distant or have differences in economic conditions, such as India, the correlation with other neighboring countries is low or even negative. Overall, the spatial effects among countries with Macau as a tourist destination reflect the geographic, cultural and economic factors.\\
\indent In addition, tourists from mainland China and Hong Kong are the most common, leading to a stronger dominant effect in calculating the correlation coefficients. However, tourists from India and Southeast Asian countries are relatively few. The sparseness of the sample may affect the stability of the correlation statistics. Such an imbalanced distribution of samples further strengthens the main tourist sources on the spatial effects. \\
\textbf{(2) Turkey dataset}\\
\indent The heat map shows that the correlation of tourists among Western European countries (e.g., Germany, France, the Netherlands, and the United Kingdom) is with a correlation coefficient close to 1, which may be related to their similar economic backgrounds, common spending power, and travel preferences. These countries' residents are more likely to vacation in Turkey during the summer or holiday season to enjoy the Mediterranean climate and rich cultural experiences. The tourist behavior of countries geographically close to Turkey, such as Russia and Bulgaria, also reflects the consistency of travel trends with Western European countries. In addition, Russia has deep historical economic and cultural ties with Turkey (e.g., energy cooperation and historical interactions). In contrast, countries that are geographically distant or culturally different, such as Iran, have lower correlations with European countries. Overall, spatial effects on tourist flows to Turkey are significantly driven by geographic, cultural, and economic factors.

\subsection{Virtual sample performance}
We assess the sample quality generated using two quantitative metrics: the Hurst exponent \citep{tzouras2015financial} and Dynamic Time Warping (DTW) \citep{yadav2018dynamic}. The Hurst exponent and DTW quantitative metrics complement each other and ensure the quality of the virtual samples. The Hurst exponent can assess whether the generated samples exhibit similar long-term dependencies as the real sample. The DTW can assess the diversity within the generated samples. While the Hurst exponent focuses on long-term pattern alignment, the DTW provides flexibility in capturing local variations. The Hurst exponent and DTW complement each other and ensure the virtual samples are similar to the real sample and exhibit sufficient diversity. The Hurst exponent and DTW are calculated as \Eqref{H} and \Eqref{DTW} respectively.
\begin{align}\label{H}
\text H = \lim_{n \to \infty} \frac{\log(\mathbb{E}[R(p)/S(p)])}{\log(p)},
\end{align}
\begin{align}\label{DTW}
\text {DTW(X, Y)} = \min_{\pi \in \mathcal{P}(m, m)} \sqrt{\sum_{(i,j) \in \pi} (x_i - y_j)^2}.
\end{align}
$R(p)$ represents the range of the cumulative deviations from the mean, and $S(p)$ represents the standard deviation. $p$ represents the period.  $\pi$ represents the warping path that aligns the two sequences, and $\mathcal{P}(n, m)$ represents all possible warping paths set. $X = (x_1, x_2, \dots, x_m)$and $Y = (y_1, y_2, \dots, y_m)$ represents two sequences.

\subsection{Predicting performance}
we evaluate the prediction performance of our model using two commonly employed metrics: Mean Absolute Error (MAE) and Mean Absolute Percentage Error (MAPE), as shown in \Eqref{MAE} and \Eqref{MAPE}. We use the improvement Rate (IR) formula to quantify the forecasting improvement between the two models. The IR represents the percentage by which the MAPE of model 1 deviates from that of model 2 and is calculated as \Eqref{IR}.
\begin{align} \label{MAE}
\text{MAE} = \frac{1}{p} \sum_{i=1}^{p} |x_i - \hat{x}_i|,
\end{align}
\begin{align} \label{MAPE}
\text{MAPE} = \frac{1}{p} \sum_{i=1}^{p} \left|\frac{x_i - \hat{x}_i}{x_i}\right|,
\end{align}
\begin{align} \label{IR}
IR = \frac{\text{MAPE(model1)} - \text{MAPE(model2)}}{\text{MAPE(model2)}} \times 100\%.
\end{align}
where $x_i$ denotes the real value, $\hat{x}_i$ the predicted value, and $p$ the prediction step.


\subsection{Benchmark models for virtual sample generation}
Five benchmark models are employed to evaluate the performance of the proposed GAN-based models in generating virtual samples. These models include DCGAN \citep{radford2015unsupervised}, TimeGAN \citep{yoon2019time}, RCGAN \citep{esteban2017real}, WGAN-GP \citep{arjovsky2017wasserstein}, and C-RNN-GAN \citep{mogren2016c}.\\
\textbf{(1) DCGAN} effectively captures complex patterns in tourism demand samples using deep convolutional neural networks. . \\
\textbf{(2) TimeGAN} combines self-supervised learning with GAN designed explicitly for time series. TimeGAN comprises four network components: embedding, recovery, generator, and discriminator, as shown in . \\
\textbf{(3) RCGAN} integrates CGAN with LSTM focusing on generating samples based on conditions. \\
\textbf{(4) WGAN-GP} improves the stability of GAN training by incorporating the Wasserstein distance and a gradient penalty mechanism. The structure of the WGAN-GP generator and discriminator is the same as DCGAN. \\
\textbf{(5) C-RNN-GAN} combines RNNs with GAN but focuses more on handling continuous time series. The C-RNN-GAN generator generates sequences by combining random noise with the previous output at each time step. The RCGAN incorporates conditional information at each time step without feedback to the next step. The C-RNN-GAN employs a bidirectional LSTM, while RCGAN uses a unidirectional LSTM and incorporates conditional features.

\subsection{ Benchmark models for predictors}
Eight models are used as benchmark models to evaluate forecasting performance. The benchmark models can be divided into non-spatiotemporal and spatiotemporal models. Non-spatiotemporal models include time series models, such as Autoregressive Integrated Moving Average (ARIMA), Exponential Triple Smoothing (ETS) and Seasonal Na?ve (SNa?ve); deep learning models, such as LSTM, the traditional Transformer and Informer. The spatiotemporal model includes the Spatiotemporal Autoregressive Combined model (STAC) \citep{jiao2021forecasting} and Spatiotemporal Fusion Graph Convolutional Network (ST-FGCN) \citep{li2022tourism}.

\subsection{Parameter settings for all models}
The experiment is divided into two parts. The first part is to compare the Hurst exponent and DTW distance of the sample generated by the proposed spatiotemporal GAN with five GAN-based models (DCGAN, TimeGAN, RCGAN, WGAN-GP, and C-RNN-GAN) to demonstrate the superiority of the proposed spatiotemporal GAN. The second part is to select the optimal four GAN-based models and the proposedpredictor for tourism demand forecasting after joint training and compare the results with eight benchmark models (ARIMA, ETS, SNa?ve, LSTM, Transformer, Informer, STAC and ST-FGCN). To verify the effectiveness of joint training and predictor, we evaluate the contribution by removing or replacing them. We replace the joint training strategy directly with the direct sample piecing method to verify the joint training, in which the predictor directly inputs the virtual sample and the real sample (called proposed model ${\dagger}$ (delete joint training)). To verify the effectiveness of the predictor, we replace the predictor with the Informer predictor and conduct joint training with the proposed virtual sample generation method (called Informer joining proposed model*).

We use the rolling window technique in our experiments. At each window, the model will forecast the next $p$ steps. The hyper parameter ranges are shown in \tabref {hyper_parameters}.

\begin{table}[H]
\centering
\setlength{\abovecaptionskip}{0cm} 
\setlength{\belowcaptionskip}{-0.2cm}
\begin{threeparttable}
\caption{Hyperparameter settings for the proposed predictor.}
\label{hyper_parameters}
\begin{tabularx}{\linewidth}{X c}
\toprule
Hyperparameter & Value \\
\midrule
Model dimension & 512 \\
Number of attention heads & 8 \\
Number of encoder layers & 6 \\
Feedforward network dimension & 2048 \\
Dropout rate & 0.1 \\
Learning rate & 2e-4 \\
Batch size & 64 \\
Kernel size of causal convolution & 3 \\
\bottomrule
\end{tabularx}
\end{threeparttable}
\end{table}

\begin{table}[H]
\centering
\setlength{\abovecaptionskip}{0cm} 
\setlength{\belowcaptionskip}{-0.2cm}
\caption{Comparison of GAN-based models using the Hurst Exponent and DTW Distance (Macau dataset).}
\label{Comparison of}
\begin{threeparttable}
\resizebox{\textwidth}{!}{
\begin{tabular}{cccccccccccccc}
\toprule
 \multirow{2}{*}{Country/Region} & \multirow{2}{*} {$H_{\text{real}}$} & \multicolumn{2}{c}{DCGAN} & \multicolumn{2}{c}{TimeGAN} & \multicolumn{2}{c}{RCGAN} & \multicolumn{2}{c}{WGAN-GP} & \multicolumn{2}{c}{C-RNN-GAN} & \multicolumn{2}{c}{Proposed model*} \\
\cmidrule(lr){3-4} \cmidrule(lr){5-6} \cmidrule(lr){7-8} \cmidrule(lr){9-10} \cmidrule(lr){11-12} \cmidrule(lr){13-14}
 &  & H & DTW & $H$ & DTW & H & DTW & $H$ & DTW & $H$ & DTW & $H$ & DTW \\
\midrule
Mainland China & 0.77 & \textcolor{copper}{0.53} & \textcolor{gold}{136.69} & 0.52 & 147.54 & 0.52 & \textcolor{copper}{153.29} & \textcolor{copper}{0.53} & \textcolor{silver}{146.60} & \textcolor{silver}{0.55} & 385.59 & \textcolor{gold}{0.56} & \textcolor{silver}{140.60} \\
Hong Kong & 0.71 & 0.49 & \textcolor{gold}{145.67} & \textcolor{copper}{0.53} & 165.54 & \textcolor{copper}{0.53} & 170.59 & 0.52 & \textcolor{silver}{150.65} & \textcolor{silver}{0.54} & 366.08 & \textcolor{gold}{0.62} & \textcolor{copper}{164.09} \\
Taiwan & 0.56 & \textcolor{copper}{0.53} & \textcolor{gold}{101.91} & \textcolor{silver}{0.55} & 158.59 & \textcolor{copper}{0.53} & \textcolor{copper}{130.51} & 0.52 & 153.46 & \textcolor{gold}{0.56} & 440.13 & \textcolor{gold}{0.56} & \textcolor{silver}{123.22} \\
Korea & 0.71 & 0.52 & \textcolor{gold}{135.31} & \textcolor{copper}{0.56} & 209.33 & 0.52 & \textcolor{silver}{139.84} & \textcolor{copper}{0.56} & 205.15 & \textcolor{copper}{0.55} & 336.66 & \textcolor{gold}{0.58} & \textcolor{copper}{159.47} \\
Japan & 0.60 & 0.52 & 290.21 & \textcolor{silver}{0.54} & \textcolor{copper}{184.67} & \textcolor{silver}{0.54} & 187.32 & \textcolor{copper}{0.53} & \textcolor{gold}{156.13} & \textcolor{silver}{0.54} & 306.08 & \textcolor{gold}{0.62} & \textcolor{silver}{167.59} \\
Philippines & 0.76 & \textcolor{copper}{0.53} & 289.65 & 0.50 & \textcolor{silver}{177.91} &  {0.51} & 253.54 & \textcolor{copper}{0.53} & \textcolor{copper}{244.07} & \textcolor{silver}{0.54} & 271.81 & \textcolor{gold}{0.63} & \textcolor{gold}{172.25} \\

Malaysia & 0.69 & \textcolor{gold}{0.60} & \textcolor{copper}{150.40} & 0.51 & {179.30} & 0.53 & \textcolor{gold}{134.27} & \textcolor{gold}{0.60} & \textcolor{silver}{140.18} & \textcolor{silver}{0.56} & 357.07 & \textcolor{silver}{0.56} & \textcolor{silver}{140.18} \\

Thailand & 0.69 & \textcolor{silver}{0.55} & \textcolor{gold}{135.64} & 0.52 & 163.55 & \textcolor{copper}{0.54} & \textcolor{silver}{140.64} & 0.54 & 172.20 & \textcolor{silver}{0.55} & 369.47 & \textcolor{gold}{0.62} & \textcolor{copper}{150.54} \\
Indonesia & 0.64 & 0.48 & \textcolor{copper}{199.18} & 0.52 & \textcolor{silver}{178.83} & 0.52 & 238.00 & \textcolor{silver}{0.56} & 236.79 & \textcolor{copper}{0.55} & 254.66 & \textcolor{gold}{0.58} & \textcolor{gold}{178.48} \\
India & 0.66 & \textcolor{gold}{0.60} & \textcolor{silver}{182.62} & 0.51 & \textcolor{copper}{187.39} & 0.49 & 210.05 & \textcolor{silver}{0.55} & 309.98 & \textcolor{copper}{0.54} & 276.22 & \textcolor{gold}{0.60} & \textcolor{gold}{164.83} \\
Singapore & 0.68 & 0.52 & \textcolor{gold}{122.28} & 0.52 & \textcolor{copper}{140.28} & \textcolor{copper}{0.53} & {235.51} & 0.50 & {157.79} & \textcolor{silver}{0.55} & 335.23 & \textcolor{gold}{0.62} & \textcolor{silver}{134.83} \\
\midrule
Average & & 0.53 & \textcolor{copper}{171.78}& 0.53 & \textcolor{silver}{170.27} & 0.52 & 181.23 & \textcolor{copper}{0.54} & 188.45 & \textcolor{silver}{0.55} & 336.27 & \textcolor{gold}{0.60} & \textcolor{gold}{154.19} \\
\bottomrule
\end{tabular}
}
\begin{tablenotes}
  \footnotesize
  \item[1] \textcolor{gold}{Gold} represents the best result; \textcolor{silver}{silver} represents the second best result; \textcolor{copper}{copper} represents the third best result.
  \item[2] Use the * symbol to mark the proposed model for virtual sample generation.
\end{tablenotes}
\end{threeparttable}
\end{table}

\begin{table}[H]
\centering
\setlength{\abovecaptionskip}{0cm} 
\setlength{\belowcaptionskip}{-0.2cm} 
\caption{Comparison of GAN-based Models using the Hurst Exponent and DTW Distance (Turkey dataset).}
\label{Comparison of 1}
\begin{threeparttable}
\resizebox{\textwidth}{!}{
\begin{tabular}{lcccccccccccccc}
\toprule
 \multirow{2}{*}{Country} & \multirow{2}{*} {$H_{\text{real}}$} & \multicolumn{2}{c}{DCGAN} & \multicolumn{2}{c}{TimeGAN} & \multicolumn{2}{c}{RCGAN} & \multicolumn{2}{c}{WGAN-GP} & \multicolumn{2}{c}{C-RNN-GAN} & \multicolumn{2}{c}{Proposed model*} \\
\cmidrule(lr){3-4} \cmidrule(lr){5-6} \cmidrule(lr){7-8} \cmidrule(lr){9-10} \cmidrule(lr){11-12} \cmidrule(lr){13-14}
 &  & H & DTW & $H$ & DTW & H & DTW & $H$ & DTW & $H$ & DTW & $H$ & DTW \\
\midrule
Germany         & 0.58 & 0.50 & 28.60 & \textcolor{silver}{0.54} & \textcolor{copper}{25.85} & 0.50 & 47.20 & 0.51 & \textcolor{silver}{24.40} & \textcolor{copper}{0.52} & 31.60 & \textcolor{gold}{0.57} & \textcolor{gold}{23.91} \\
Bulgaria        & 0.73 & 0.53 & 38.56 & \textcolor{copper}{0.54} & \textcolor{silver}{24.52} & 0.50 & \textcolor{copper}{34.25} & \textcolor{gold}{0.59} & 24.87 & 0.51 & 35.06 & \textcolor{silver}{0.58} & \textcolor{gold}{22.59} \\
France          & 0.63 & 0.51 & \textcolor{copper}{27.24} &{0.54} & 33.08 & \textcolor{silver}{0.58} & \textcolor{gold}{21.27} & \textcolor{copper}{0.55}  & 30.35 & 0.51 & 39.38 & \textcolor{gold}{0.59} & \textcolor{silver}{26.33} \\
Netherlands     & 0.58 & \textcolor{silver}{0.56} & 36.09 & \textcolor{copper}{0.54} & \textcolor{copper}{30.05} & 0.52 & \textcolor{silver}{27.20} & \textcolor{copper}{0.54} & 31.25 & \textcolor{copper}{0.54} & 38.13 & \textcolor{gold}{0.59} & \textcolor{gold}{24.30} \\
United Kingdom  & 0.57 & 0.52 & \textcolor{copper}{32.80} & \textcolor{copper}{0.53} & 33.90 & 0.49 & \textcolor{gold}{31.20} & 0.53 & 34.29 & \textcolor{silver}{0.55} & 35.00 & \textcolor{gold}{0.60} & \textcolor{silver}{31.90} \\
U.S.A           & 0.72 & \textcolor{copper}{0.53} & 37.35 & \textcolor{silver}{0.54} & \textcolor{silver}{25.30} & \textcolor{copper}{0.53} & \textcolor{gold}{25.12} & 0.50 & \textcolor{copper}{26.44} & 0.52 & 32.15 & \textcolor{gold}{0.60} & {26.96} \\
Azerbaijan      & 0.71 & \textcolor{gold}{0.57} & 39.63 & \textcolor{copper}{0.54} & \textcolor{copper}{31.43} & 0.51 & \textcolor{silver}{29.87} & \textcolor{silver}{0.55} & \textcolor{gold}{19.04} & 0.53 & 46.95 & \textcolor{gold}{0.57} & {31.52} \\
Moldova         & 0.66 & 0.52 & 40.87 & \textcolor{silver}{0.55} & 35.20 & 0.50 & \textcolor{silver}{23.50} & 0.55 & \textcolor{copper}{34.43} & \textcolor{copper}{0.54} & 41.36 & \textcolor{gold}{0.59} & \textcolor{gold}{20.48} \\
Iran            & 0.69 & \textcolor{copper}{0.54} & 38.29 & \textcolor{silver}{0.55} & \textcolor{copper}{30.50} & 0.49 & 35.24 & 0.53 & \textcolor{silver}{29.60} & \textcolor{silver}{0.55} & 41.91 & \textcolor{gold}{0.57} & \textcolor{gold}{28.73} \\
Russia          & 0.63 & \textcolor{copper}{0.53} & \textcolor{silver}{30.09} & {0.52} &  \textcolor{copper}{32.48} & {0.52} & {34.85} & \textcolor{silver}{0.54} & 39.68 & 0.45 & 47.55 & \textcolor{gold}{0.56} & \textcolor{gold}{24.85} \\
\midrule
Average         & 0.65 & \textcolor{copper}{0.53} & 34.95 & \textcolor{silver}{0.54} & \textcolor{copper}{30.23} & {0.51} & {30.97} & \textcolor{silver}{0.54} & \textcolor{silver}{29.44} & {0.52} & 38.91 & \textcolor{gold}{0.58} & \textcolor{gold}{26.16} \\
\bottomrule
\end{tabular}}
\begin{tablenotes}
  \footnotesize
  \item[1] \textcolor{gold}{Gold} represents the best result; \textcolor{silver}{silver} represents the second best result; \textcolor{copper}{copper} represents the third best result.
  \item[2] Use the * symbol to mark the proposed model for virtual sample generation.
\end{tablenotes}
\end{threeparttable}
\end{table}

\begin{table}[htb]
\centering
\setlength{\abovecaptionskip}{0cm} 
\setlength{\belowcaptionskip}{-0.2cm} 
\caption{The average performance of all models at 1, 3, 5 and 14 steps ahead (Macau dataset).}
\label{The average performance}
\begin{threeparttable}
\begin{tabularx}{\textwidth}{p{1.5cm}*{8}{X}}
\hline
\multirow{2}{*}{Model}
& \multicolumn{2}{c}{1-step}
& \multicolumn{2}{c}{3-steps}
& \multicolumn{2}{c}{5-steps}
& \multicolumn{2}{c}{14-steps} \\
\cmidrule(r){2-3} \cmidrule(r){4-5} \cmidrule(r){6-7} \cmidrule(r){8-9}
& MAE & MAPE & MAE & MAPE & MAE & MAPE & MAE & MAPE \\
\midrule
M1   & 2321.50 & 0.216 & 2341.72 & 0.329 & 2522.48 & 0.308 & 3115.84 & 0.399 \\
M2   & 1778.58 & 0.163 & 2145.84 & 0.274 & 2276.26 & 0.288 & 2833.68 & 0.342 \\
M3   & 1506.94 & 0.138 & 1627.25 & 0.171 & 2215.16 & 0.224 & 2850.13 & 0.307 \\
M4   & 1861.46 & 0.181 & 2310.50 & 0.282 & 2585.36 & 0.329 & 3040.88 & 0.415 \\
M5   & 1647.78 & 0.152 & 2362.30 & 0.237 & 2494.18 & 0.300 & 2665.52 & 0.337 \\
M6   & 1515.83 & 0.139 & 1735.56 & 0.212 & 2059.48 & 0.218 & 2606.60 & 0.286 \\
M7   & 1525.10 & 0.123 & 1713.17 & 0.140 & 1982.88 & 0.161 & 2305.51 & 0.189 \\
M8   & 1449.57 & 0.120 & 1655.56 & 0.134 & 1833.75 & 0.151 & 2295.39 & 0.179 \\
M9   & 1338.72 & 0.120 & 1671.24 & 0.137 & 1938.94 & 0.144 & 2308.09 & 0.165 \\
M10  & 1319.45 & 0.118 & 1590.42 & 0.128 & 1893.60 & 0.158 & 2014.32 & 0.166 \\
M11  & 1125.95 & 0.109 & 1507.18 & 0.116 & 1693.52 & 0.129 & 1980.93 & 0.147 \\
M12  & 1025.80 & \textbf{0.094} & 1438.63 & \textbf{0.108} & 1520.96 & \textbf{0.119} & 1720.13 & \textbf{0.132} \\
M13  & 1265.54 & 0.116 & 1496.68 & 0.131 & 1823.06 & 0.141 & 2263.69 & 0.163 \\
M14  & 1199.48 & 0.107 & 1312.27 & 0.120 & 1615.34 & 0.132 & 1990.50 & 0.146 \\
\hline
\end{tabularx}
\begin{tablenotes}
  \footnotesize
  \item[1] \parbox[t]{0.95\textwidth}{M1 = ARIMA, M2 = SNa\"ive, M3 = ETS, M4 = LSTM, M5 = Transformer, M6 = Informer, M7 = STAC, M8 = ST-FGCN, M9 = predictor joining DCGAN, M10 = predictor joining TimeGAN, M11 = predictor joining WGAN-GP, M12 = Ours , M13 = Ours (delete joint training), M14 = Informer joining proposed model*.}
  \item[2] \textbf{Bold} indicates the best MAPE for each step.
\end{tablenotes}
\end{threeparttable}
\end{table}

\begin{table}[H]
\centering
\setlength{\abovecaptionskip}{0cm} 
\setlength{\belowcaptionskip}{-0.5cm} 
\caption{The average performance of all models at 1, 2, 6 and 12 steps ahead (Turkey dataset).}
\label{The average performance 1}
\begin{threeparttable}
\begin{tabularx}{\textwidth}{p{1.5cm}*{8}{X}}
\toprule
\multirow{2}{*}{Model}
& \multicolumn{2}{c}{1-step}
& \multicolumn{2}{c}{3-steps}
& \multicolumn{2}{c}{6-steps}
& \multicolumn{2}{c}{12-steps} \\
\cmidrule(r){2-3} \cmidrule(r){4-5} \cmidrule(r){6-7} \cmidrule(r){8-9}
& MAE & MAPE & MAE & MAPE & MAE & MAPE & MAE & MAPE \\
\midrule
M1   & 56579.76 & 0.206 & 66350.18 & 0.329 & 110972.30 & 0.301 & 99927.88 & 0.395 \\
M2   & 122238.50 & 0.167 & 99259.67 & 0.248 & 92728.42 & 0.282 & 222722.52 & 0.324 \\
M3   & 61905.31 & 0.135 & 97218.88 & 0.171 & 91920.84 & 0.236 & 244087.57 & 0.285 \\
M4   & 50697.88 & 0.155 & 60616.85 & 0.218 & 65996.88 & 0.245 & 58346.97 & 0.301 \\
M5   & 40771.68 & 0.156 & 54293.09 & 0.227 & 69588.67 & 0.295 & 75329.70 & 0.342 \\
M6   & 88381.08 & 0.161 & 58984.33 & 0.203 & 54971.18 & 0.247 & 58770.52 & 0.312 \\
M7   & 37301.58 & 0.121 & 47555.98 & 0.145 & 37946.71 & 0.190 & 39939.42 & 0.231 \\
M8   & 71455.00 & 0.117 & 92592.96 & 0.131 & 72035.64 & 0.166 & 83134.59 & 0.205 \\
M9   & 43805.90 & 0.123 & 41808.36 & 0.152 & 42159.72 & 0.168 & 39184.10 & 0.201 \\
M10  & 42444.69 & 0.121 & 42907.63 & 0.131 & 41545.11 & 0.150 & 39869.40 & 0.185 \\
M11  & 49118.71 & 0.119 & 55152.16 & 0.128 & 52666.97 & 0.137 & 49064.56 & 0.162 \\
M12  & 39151.67 & \textbf{0.103} & 51442.42 & \textbf{0.120} & 48488.05 & \textbf{0.128} & 32308.46 & \textbf{0.151} \\
M13  & 53044.98 & 0.115 & 47334.96 & 0.126 & 44380.48 & 0.134 & 61263.22 & 0.164 \\
M14  & 47334.96 & 0.115 & 59488.31 & 0.123 & 59488.31 & 0.131 & 44380.48 & 0.161 \\
\bottomrule
\end{tabularx}
\begin{tablenotes}
  \footnotesize
  \item[1] \parbox[t]{0.95\textwidth}{M1 = ARIMA, M2 = SNa?ve, M3 = ETS, M4 = LSTM, M5 = Transformer, M6 = Informer, M7 = STAC, M8 = ST-FGCN, M9 = predictor joining DCGAN, M10 = predictor joining TimeGAN, M11 = predictor joining WGAN-GP, M12 =Ours, M13 = Ours (delete joint training), M14 = Informer joining proposed model*.}
  \item[2] \textbf{Bold} indicates the best MAPE for each step.
\end{tablenotes}
\end{threeparttable}
\end{table}

\subsection{Virtual sample generation results analysis}
The virtual sample quality is judged by calculating the comparative results of the Hurst exponent and the DTW distance, as shown in \tabref{Comparison of} and \tabref{Comparison of 1}. We mark each model's best, second-best, and third-best performance in different regions in gold, silver, and copper colors.
The \tabref{Comparison of} and \tabref{Comparison of 1} show that the proposed method outperforms other models on both the Macau and Turkey datasets. DCGAN captures high-frequency variations better on DTW due to its simple structure. However, DCGAN has a low Hurst exponent, suggesting a challenge of long-term dependence. TimeGAN enhances the smoothness modeling of the time trend by using the GRU module, and its Hurst exponent performs smoothly. RCGAN  improves performance through conditional labels, but it is less effective than TimeGAN. The performance of WGAN-GP is attributed to the optimization of the gradient penalty term and the Wasserstein distance when compared to the DCGAN structure. WGAN-GP is better at capturing long-time series features. C-RNN-GAN performs average and smoothly overall. In the Macau dataset, the generated samples are very close to the original samples in long-term dependencies, but the DTW distance is large. Overall, the proposed method achieves high smoothness and dynamic diversity of the generated samples on both datasets. In particular, the proposed method performs the best by combining the advantages of DCGAN, TimeGAN, and WGAN-GP.

\subsection{The predictor results analysis}
\begin{figure}[H]\centering
       \includegraphics[width=\linewidth]{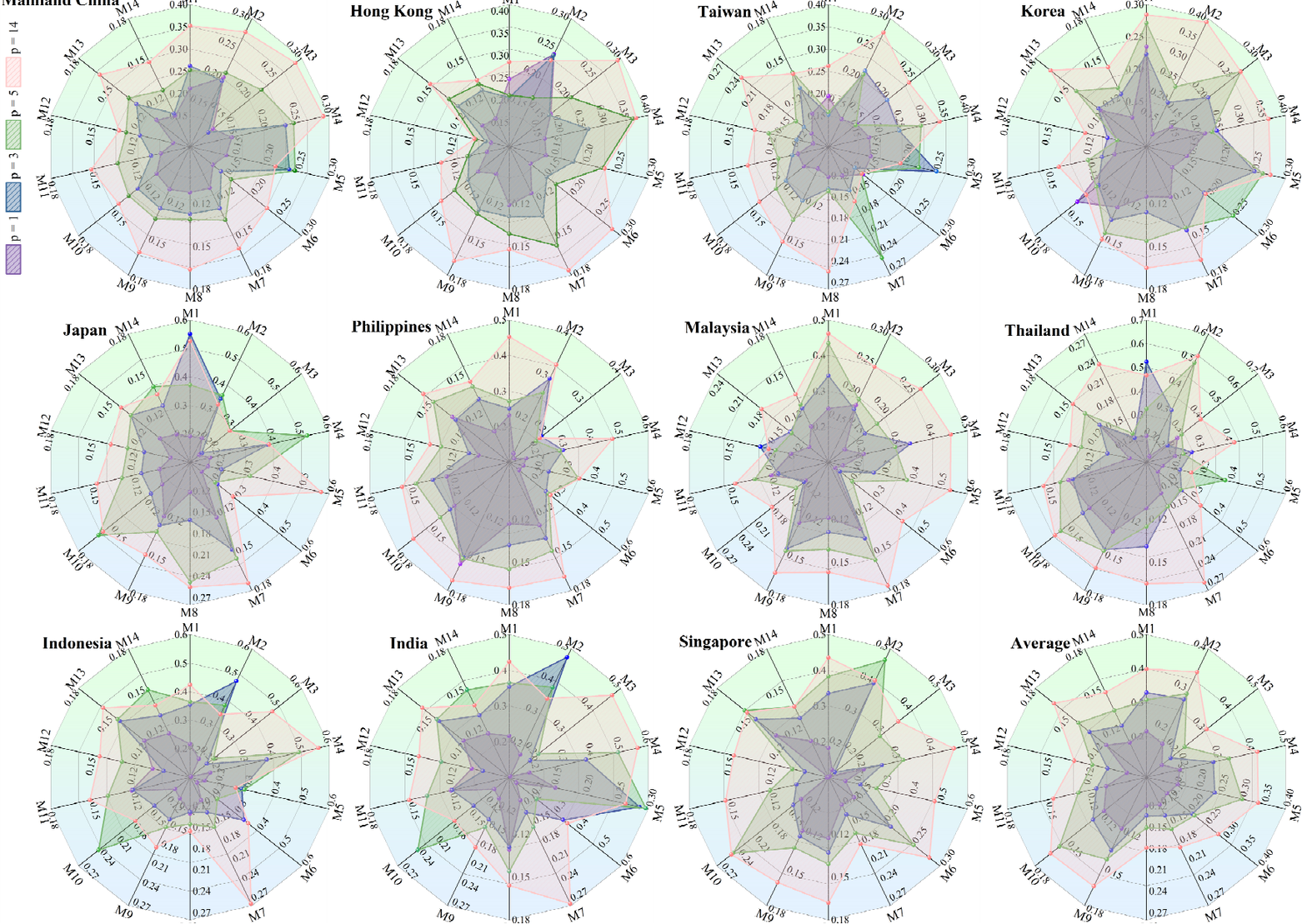}
	\caption{Radar chart of MAPE for tourism demand using various models (Macau dataset: purple, blue, green, and pink were used to fill the MAPE forecasting horizons of 1, 3, 5, and 14 steps).}
    \label{Radar chart}
\end{figure}

\begin{figure}[H]\centering
       \includegraphics[width=\linewidth]{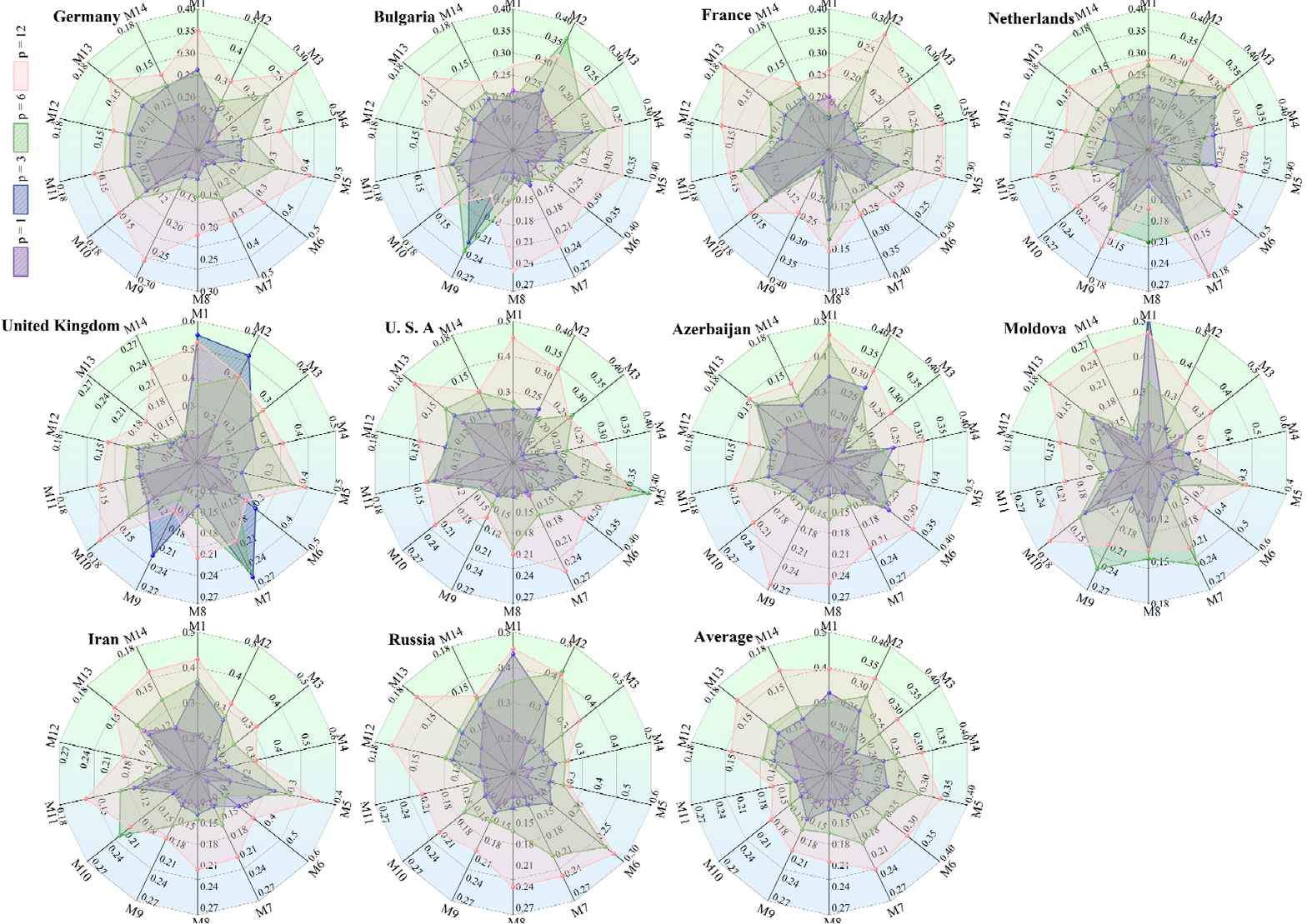}
	\caption{Radar chart of MAPE for tourism demand using various models (Turkey dataset: purple, blue, green, and pink were used to fill the MAPE forecasting horizons of 1,3, 6, and 12 steps).}
    \label{Radar chart 1}
\end{figure}

The experiment is divided into three parts. The first part trains the benchmark models (ARIMA, ETS, SNa?ve, LSTM, Transformer, Informer, STAC and ST- FGCN) for tourism demand forecasting. The second part is the joint training of the enhanced proposed predictor and GAN-based models (predictor joining DCGAN, predictor joining TimeGAN, predictor joining WGAN-GP and Ours). In the third part, we train the ablation model (Ours (delete joint training), Informer joining proposed model*) and compare The average performance of the proposed model with the benchmark and joint training models. The average performance of the Macau and Turkey datasets are shown in \tabref{The average performance} and \tabref{The average performance 1}. M1 to M14 represent ARIMA, SNa?ve, ETS, LSTM, Transformer, Informer, STAC, ST-FGCN, predictor joining DCGAN, predictor joining TimeGAN, predictor joining WGAN-GP, Ours, Ours (delete joint training), Informer joining proposed model*. In addition, \tabref{The average performance} and \tabref{The average performance 1} indicate that the proposed model consistently achieves the lowest error across most forecast horizons. We draw a radar map of MAPE for each country or region across all forecast horizons as shown in \figref{Radar chart} and \figref{Radar chart 1}. The radar charts visualize the MAPE values of the various models over different prediction horizons and further confirm the robustness of the proposed model.

Time series models, such as the SNa?ve and ETS models, perform better on average, as shown in \tabref{The average performance} and \tabref{The average performance 1}. SNa?ve has a much lower MAPE than ARIMA by replicating historical seasonal patterns. ETS consistently outperforms the other time series models in the Macau and Turkey datasets in multiple horizon forecasts. The reason is the ETS's pre-determined trend and seasonal structure. In particular, the MAPE of ETS is lower than that of Transformer in the 1, 3-step ahead and 5,14-step ahead forecasts in the Macau dataset. ETS is better than Informer in 1 and 3-step ahead. ETS outperforms mostly Transformer and Informer in the Turkey dataset. The experimental results for ETS illustrate two factors. First, the performance of simple models can be excellent. The performance of a model is not proportional to the complexity of the model. Second, Transformer and Informer contain multiple attention heads and layers of stacked encoders and decoders. Each layer contains a huge number of trainable parameters. Limited samples are not enough to train large parameters.

\tabref{The average performance} and \tabref{The average performance 1} demonstrate that spatiotemporal models consistently outperform non-spatiotemporal models. As a representative of non-spatiotemporal models, the ETS model works well for a wide range of prediction horizons. However, ETS is less effective than the spatiotemporal models STAC and ST-FGCN. The spatial effects are considered in STAC and ST-FGCN to learn the sample features better. The STAC model overcomes the limitations of the spatial autoregressive combined model in handling spatiotemporal dynamics. The FGCN model introduces a GCN, and the spatial weight matrix incorporates various information to capture spatial features, such as geographic proximity and demand similarity. Although the STAC model can dynamically consider spatial relationships, it is less effective than ST-FGCN in processing nonlinear relationships.

We propose the enhanced Transformer predictor trained jointly with the spatiotemporal GAN, and the Transformer's prediction is greatly improved as shown in \tabref{The average performance} and \tabref{The average performance 1}. In both the Macau and Turkey datasets, the MAPE of the proposed Transformer is always lower than that of other spatiotemporal models in multiple prediction horizons. This stems from the enhanced predictor's ability to dynamically learn spatiotemporal features using sufficient data, which combines STAC and ST-FGCN's advantages. In addition, we jointly train the enhanced Transformer with the DCGAN, TimeGAN, and WGAN-GP. However, the effect is not as good as the joint training of the spatiotemporal GAN with the predictor. The joint training experiments demonstrate that the quality of data generated by our proposed model outperforms other GAN-based models.

\begin{table}[H]
\centering
\setlength{\abovecaptionskip}{0cm}
\setlength{\belowcaptionskip}{-0.2cm}
\begin{threeparttable}
{\caption{The average improvement rate of all forecasting horizons between different models (Macau dataset).}
\label{Improvement Rates}
\begin{tabularx}{\textwidth}{p{9cm}*{2}{X}}
\toprule
Comparison & IR(MAE) & IR(MAPE) \\
\midrule
Ours vs ARIMA & 44.61 & 63.78 \\
Ours vs SNa?ve & 36.85 & 57.48 \\
Ours vs ETS & 30.42 & 46.00 \\
Ours vs LSTM & 41.77 & 62.43 \\
Ours vs Transformer & 37.78 & 55.76 \\
Ours vs Informer & 27.94 & 46.85 \\
Ours vs STAC & 24.20 & 25.88 \\
Ours vs ST-FGCN & 21.13 & 22.23 \\
Ours vs predictor joining DCGAN & 21.38 & 19.82 \\
Ours vs predictor joining TimeGAN & 16.31 & 20.37 \\
Ours vs predictor joining WGAN-GP & 9.55 & 9.30 \\
Ours vs Ours (delete joint training) & 16.70 & 17.69 \\
Ours vs Informer joining proposed model* & 6.74 & 9.86 \\
predictor joining DCGAN vs Transformer & 20.86 & 44.82 \\
predictor joining TimeGAN vs Transformer & 25.65 & 44.44 \\
predictor joining WGAN-GP vs Transformer & 31.21 & 51.22 \\
Ours (delete joint training) vs Transformer & 25.31 & 46.25 \\
\bottomrule
\end{tabularx}
}
\end{threeparttable}
\end{table}

\begin{table}[H]
\centering
\setlength{\abovecaptionskip}{0cm}
\setlength{\belowcaptionskip}{-0.5cm}
\begin{threeparttable}
\caption{The average improvement rate of all forecasting horizons between different models (Turkey dataset).}
\label{Improvement Rates 1}
\begin{tabularx}{\textwidth}{p{9cm}*{2}{X}}
\toprule
Comparison & IR(MAE) & IR(MAPE) \\
\midrule
Ours vs ARIMA & 48.66 & 59.27 \\
Ours vs SNaive & 68.08 & 50.84 \\
Ours vs ETS & 65.38 & 39.32 \\
Ours vs LSTM & 27.27 & 45.40 \\
Ours vs Transformer & 28.58 & 50.83 \\
Ours vs Informer & 34.36 & 45.69 \\
Ours vs STAC & 5.31 & 27.02 \\
Ours vs ST-FGCN & 46.31 & 18.63 \\
Ours vs predictor joining DCGAN & 2.65 & 22.19 \\
Ours vs predictor joining TimeGAN & 2.77 & 14.45 \\
Ours vs predictor joining WGAN-GP & 16.80 & 8.10 \\
Ours vs Ours (delete joint training) & 16.81 & 6.91 \\
Ours vs Informer joining proposed model* & 18.65 & 5.18 \\
predictor joining DCGAN vs Transformer & 30.43 & 36.81 \\
predictor joining TimeGAN vs Transformer & 30.51 & 42.53 \\
predictor joining WGAN-GP vs Transformer & 14.16 & 46.50 \\
Ours (delete joint training) vs Transformer & 14.15 & 47.19 \\
\bottomrule
\end{tabularx}
\end{threeparttable}
\end{table}

To reflect the improvement rate of the proposed model compared with other models, we compare the average improvement rate of the proposed model with the benchmark models and the jointly trained model, as shown in \tabref{Improvement Rates} and \tabref{Improvement Rates 1}. The improvement rates demonstrate the proposed model exhibits consistently higher average improvement rates across most forecasting horizons.
First, the proposed model improves the MAPE by 63.78\%, 57.48\% and 46.00\% on the Macau dataset compared to ARIMA, SNa?ve and ETS, respectively; and the corresponding improvement rates are 59.27\%, 50.84\% and 39.32\% on the Turkey dataset. These results show that our method can significantly improve the forecasting accuracy and show consistent improvement in datasets with different geographic features. Compared with Transformer, the proposed model improves 55.76\% and 50.83\% on the Macau and Turkey datasets, respectively, which suggests that the strategy of virtual sample generation and joint training provides the model with an advantage in capturing spatiotemporal features.

\tabref{Improvement Rates} and \tabref{Improvement Rates 1} are used to further analyze the spatiotemporal model. TThe proposed model improves the MAPE on the Macau dataset by 25.88\% and 22.23\% compared to STAC and ST-FGCN, respectively; on the Turkey dataset, this improvement is 27.02\% and 18.63\%, respectively. The proposed model can focus on the nonlinearity of the samples and employ adaptive dynamics to learn the spatial features of the samples. In the joint training approach, the proposed model improves the MAPE on the Macau dataset by 19.82\%, 20.37\% and 9.30\%, respectively, and on the Turkey dataset by 22.19\%, 14.45\% and 8.10\% correspondingly when compared to the transformer joint training using DCGAN, TimeGAN and WGAN-GP models. These results show that using different virtual sample generation methods to train the predictor jointly improves the prediction, but not as much as our proposed spatiotemporal GAN. The results suggest that the more similar the virtual samples are to the real samples, the better the prediction. The effectiveness of spatial feature methods in dynamic learning is further validated.

After canceling the joint training strategy, the MAPE improvement rates of the proposed model on the Macau and Turkey datasets are 17.69\% and 6.91\%, respectively, which further proves the importance of synchronized evolution of the generator and predictor and real-time tuning of the parameters. In addition, when using the Informer predictor instead of the Transformer, the joint model has a MAPE improvement of 9.86\% and 5.18\%. This suggests that the enhanced Transformer outperforms the Informer in capturing interaction features.

\subsection{Sensitivity and reliability assessments}

We analyze the performance of the models under different forecasting steps and time granularity using Macau and Turkey datasets. In the experiments, we use a rolling window approach containing 90 days of data to capture the short-term patterns in daily tourist arrivals in the Macau dataset. In contrast, the rolling window for the Turkey dataset is 12 months long. Each window contains a year's worth of monthly data reflecting seasonal and long-term tourism demand trends. We use the Macau dataset to forecast tourist arrivals for 1, 3, 5 and 14 days, focusing on evaluating the model's performance in short-term forecasting. In addition, we use the Turkey dataset to forecast tourism demand for 1 month, 3 months (quarterly), 6 months (semi-annually) and 12 months (annually), which validate the model's performance in medium- and long-term forecasting scenarios.\\
\indent The experimental results show in \tabref{The average performance} and \tabref{The average performance 1} that the proposed model is susceptible to short-step prediction in the Macau dataset, with MAPEs of 0.094 and 0.108 for the 1-step and 3-step predictions, respectively, which are much better than other models and can accurately capture the fast-varying trends in the daily data. The MAPE of the Macau tourism demand forecasts in the 5-step and 14-step forecasts are still better than the benchmark model, although the MAPE of the forecasts slightly increases to 0.119 and 0.132. In the Turkey demand forecast, the MAPE in the 1-step and quarterly forecasts is 0.103 and 0.120, showing its ability to capture short-term and quarterly demand changes accurately. In the half-yearly and annual step forecasts, the model outperforms the benchmark model despite the MAPE rising to 0.128 and 0.151.\\
\indent The experimental results reflect the model's ability to sensitively capture the rapid changes in the daily data in the Macau dataset. In the Turkey dataset, the model accurately identifies long-term trends and seasonal fluctuations in the monthly data. This complementarity in temporal granularity not only enhances the comprehensiveness of the experiments but also further validates the model's broad adaptability and sensitivity in complex scenarios. Therefore, we validate the robustness and adaptability of the proposed model with experimental results under different datasets, forecasting steps and time granularities.

\section{Conclusion}
\label{Conclusion}

\subsection{Managerial Implications}
The proposed forecasting framework provides valuable insights for tourism management, particularly in dynamic and uncertain environments. The model enables more effective resource allocation and crisis management decision-making by generating virtual samples that simulate realistic tourism demand patterns. For instance, during sudden disruptions such as a pandemic, the model can forecast demand shifts across different regions, allowing tourism authorities to adjust capacity, transportation, and staffing in real-time. Moreover, for daily operations and peak-season planning, capturing dynamic spatiotemporal relationships facilitates optimal tourism resource scheduling, such as accommodation availability, attraction capacity management, and travel route adjustments. This approach empowers stakeholders to adopt proactive, data-driven strategies, enhancing resilience and operational efficiency in the tourism sector.

\subsection{Research Findings and Future Work}
This study proposes a novel forecasting framework to address the challenges of sample scarcity and dynamic spatial features in spatiotemporal tourism demand forecasting. The proposed model integrates the spatiotemporal GAN with an enhanced Transformer predictor to enhance forecasting accuracy. The spatiotemporal GAN is used to generate virtual samples. The generator and discriminator of the spatiotemporal GAN use GCN with memory mechanisms to capture spatial and temporal dependencies. Furthermore, GCN can dynamically update the spatial weight matrix due to a rolling window, which helps the model learn complex and dynamic spatial features. With augmented data, the enhanced Transformer-based predictor incorporates causal convolution to capture short-term dependencies and global pooling to replace autoregressive decoding. These modifications enhance both local and global feature learning, making the model more efficient and interpretable compared to traditional Transformers.\\
\indent A joint training strategy is proposed to enhance the prediction accuracy further. This strategy embeds the Transformer's training within the spatiotemporal GAN training process, which helps the Transformer adjust models in real-time. The training process is divided into two parts: (1) pre-training the Transformer with real samples to learn fundamental patterns and establish a foundation for subsequent joint training; (2) fine-tuning the model with a combination of real samples and real-time generated virtual samples, using back propagation from the generator's loss function to regulate the Transformer's learning process.\\
\indent The experiments show the proposed spatiotemporal GAN generates virtual samples with the best Hurst Exponent and DTW distance compared with five non-spatiotemporal virtual sample generation methods. The spatiotemporal GAN generates samples with feature distributions more similar to the real samples than the model that does not consider spatiotemporal features. In addition, we select the best four virtual sample generation methods for joint training with the transformer predictor to validate the proposed methods' validity further. The experiments show that the more similar the generated virtual samples are to the real samples, the better the predictive effect of the predictor. Compared to eight benchmark models and three joint training models, the proposed predictor achieves lower prediction errors for spatiotemporal tourism demand forecasting.\\
\indent We conduct partial ablation experiments to quantify the effects of joint training and the predictor separately. Instead of joint training, we use a direct combination of virtual and real samples to train the predictor. The predictor without joint training reduces the effect by 17.69\%  and 6.91\% in Macau and Turkey datasets in MAPE. We replace the transformer predictor with Informer and the experiment shows that the effect of joint training Informer is reduced by 9.86\% and 5.18\% in MAPE compared to the proposed predictor.\\
\indent The models' superiority does not hide their limitations. The joint training phase requires substantial computational resources and careful tuning to balance the synergy between the generator and the Transformer. In the future, we will reduce training time and computational resource consumption by incorporating more efficient optimization algorithms and strategies, such as adaptive learning rates and parallel computing techniques. Additionally, we will explore more sophisticated feature extraction modules to capture better nuanced spatiotemporal relationships, such as weighted feature fusion and interaction mechanisms. Moreover, we will evaluate the model's scalability and adaptability to other domains beyond tourism demand forecasting in future work.


\appendix



\bibliographystyle{elsarticle-harv}







\end{document}